\documentclass[prd,twocolumn,tightenlines,showpacs,nofootinbib,amsfonts,amssymb,amsmath]{revtex4}
\usepackage{graphicx}
\begin{document}
\newcommand{\etal}{{\it et al.}}
\newcommand{\bx}{{\bf x}}
\newcommand{\bn}{{\bf n}}
\newcommand{\bk}{{\bf k}}
\newcommand{\dd}{{\rm d}}
\newcommand{\dslash}{D\!\!\!\!/}
\def\ga{\mathrel{\raise.3ex\hbox{$>$\kern-.75em\lower1ex\hbox{$\sim$}}}}
\def\la{\mathrel{\raise.3ex\hbox{$<$\kern-.75em\lower1ex\hbox{$\sim$}}}}
\def\beq{\begin{equation}}
\def\eeq{\end{equation}}

\vskip-4cm
\title{The Catenary Revisited: From Newtonian Strings to Superstrings}

\author{David de Klerk$^{1}$, Jeff Murugan$^{1,3}$
and Jean-Philippe Uzan$^{2,1,3}$}

\affiliation{
${^1}$ Astrophysics, Cosmology \& Gravity Center,\\ 
Department of Mathematics and Applied Mathematics\\
University of Cape Town, Rondebosch 7701 (South Africa)\\
$^2$ Institut d'Astrophysique de Paris,
         UMR-7095 du CNRS, Universit\'e Pierre et Marie
              Curie, 98 bis bd Arago, 75014 Paris (France)\\ 
${^3}$ National Institute for Theoretical Physics (NITheP),
Stellenbosch 7600 (South Africa). \\              
}
\vspace*{2cm}
\begin{abstract}
The dynamics of extended objects, such as strings and membranes, has attracted more attention
in the past decades since the fundamental objects introduced in high-energy physics
are no more pointlike. Their motion is generally intricate to describe. This article argues
that the Newtonian analogy of a catenary with free ends offers a good description
of some processes such as gravitational radiation by an accelerated brane. 
\end{abstract}
 \date{November 2010}
 \maketitle
\section{Introduction}
Much of the beauty of physics derives from the fact that it is often possible to
understand a broad range of quite complex phenomena in terms of a small class of  
simple and sometimes surprisingly universal physical problems. The harmonic oscillator, with its ubiquitous appearance everywhere from physics of diatomic molecules to the simple pendulum to the theory of cosmological perturbations, is perhaps the first example that comes to mind. As we will argue in this article, a simple suspended string in a gravitational field is another. \\

Like the harmonic oscillator, the shape of a``hanging chain" is a centuries old problem that has occupied some of the greatest mathematical minds of all time including no less than Galileo (who got it wrong), Huygens (who showed why), Leibnitz (who got it right), the Bernoulli brothers (who also got it right), Euler and Lagrange (who showed how the, then, newly developed machinery of the calculus of variations could be brought to bear on the problem with astounding efficiency). Today, not only does almost every (physics) undergraduate know that a uniform ideal chain hanging under its own weight from two fixed points at a fixed height assumes the shape of a {\it catenary}, but the determination of the shape function of the chain has become a routine exercise in variational calculus~\cite{general,variation}. So much so that one might be forgiven for thinking that the problem had nothing more of value to offer. One would, of course, be mistaken. 
Indeed, several variations on the catenary problem exist. These include the elastic, centrally loaded chains~\cite{Behroozi} and a number of studies of chains with varying mass density $\mu(s)$~\cite{Fallis}. The common theme to all of these problems though, is that they are concerned with the shape assumed by a {\it static} chain in a gravitational field. The case of a chain with {\it time-dependent} boundary conditions\footnote{Here we do not include ``towed chain" type configurations.} appears to have remained largely unstudied, as far as we are aware.\\

Our own interest in this problem arose in a largely tangential way through the study of certain instabilities of D-branes in superstring theory~\cite{conf}. D-branes are wonderfully multi-faceted membrane-like solutions in string theory that, for the purposes of this article, can be understood simply as somewhere that open strings must end. In other words, they define the boundary conditions for open strings. As such, the endpoints of the attached open string satisfy Neumann boundary conditions in the direction of the membrane\footnote{Actually we are abusing terminology a little here in the interests of intuition. ``Membrane" in string theory is usually reserved for a (2+1)-dimensional D-brane or D2-brane. We will not be fussy about the dimensionality of the extended object.} and Dirichlet conditions in the orthogonal directions. But D-branes are so much more than that; they are actually charged objects in the spectrum of the theory. They couple to Ramond-Ramond (RR) fluxes in the same way that electrons - charged particles in electromagnetism - couple to the electromagnetic field. And, in fact, just like electrons in an electromagnetic field, D-branes moving in an RR background will experience the analogue of a Lorentz force and accelerate. This is true in particular for a D-brane falling in a gravitational potential (near a black hole, for example) which will be accelerated off geodesic paths due to this Lorentz-like acceleration. Open strings attached to such D-branes, however, feel no such force since they are not RR-charged and so should be in geodesic freefall. In Ref.~\cite{Berenstein-et-al} it was argued that the effect of this difference in the way that strings and membranes behave is two-fold: the string grows and its endpoints start to come together. If the string is allowed to self interact then the open string ends coming together forms a closed string which, no longer bound to the membrane by Dirichlet boundary conditions, is free to leave, carrying away energy from the D-brane. The authors of \cite{Berenstein-et-al} discovered this instability through a sophisticated study of a class of operators that represent the membrane-string configuration in a {\it dual} quantum field theory (a supersymmetric Yang-Mills gauge theory) and, in passing, suggested that a good toy model with which to build intuition about this problem might be classical strings in Rindler space\footnote{which is the rest frame of the accelerated membrane.} or, through the equivalence principle, classical strings in a gravitational field. Some of this intuition was validated in \cite{Berenstein-Chung}, where it was demonstrated that the profile function of a (fundamental, relativistic) string suspended between two points\footnote{In string theory jargon, these would be two D0-branes which are 0-dimensional solitonic objects in the theory.} in Rindler space (a fixed distance away from the Rindler horizon) is indeed of the catenary 
form. To fully appreciate the mechanism of the D-brane instability however, we need to understand the dynamics of open strings whose endpoints are free to move along the membrane. Toward this end, and to push the analogy with the hanging chain as far as we can, we study the simple mechanical problem of an inextensible chain attached to a rod which is itself fixed at some height, h, in a gravitational field, g. The key feature to our problem is that the points of attachment are free to move along the rod (as they would be if the chain were attached by, say, frictionless rings). We would like to know
\begin{itemize}
  \item What is the shape of the string at any time, $t>0$?
  \item Intuitively, it seems clear to us that as the string 
  falls in the gravitational field, the ends come together. 
  How then does the time that it takes depend on the 
  various parameters?
  \item Does the motion of the string generate wave 
   modes along the string, and if so, what is it's frequency 
   spectrum?
\end{itemize}
To the best of our knowledge (and much to our surprise) these questions have not yet been answered for this system in the literature. In what follows, we present a detailed analysis of the dynamical hanging chain problem.\\

\section{The static catenary}
\begin{figure}[!htb]
\includegraphics[width=8cm,angle=0]{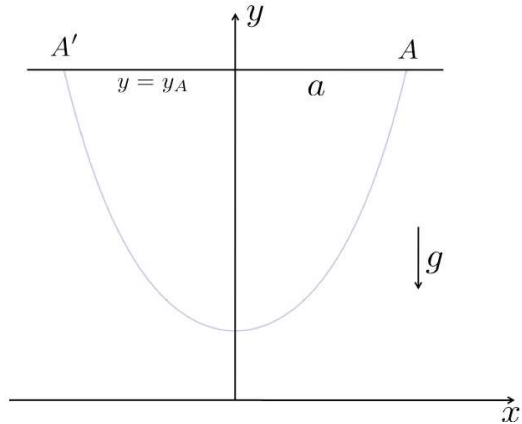}
\caption{A string of length $2\ell_0$ and lineic mass $\mu$ is hanging in a gravitational
field $g$.} 
\label{fig1}
\end{figure}

We begin with a quick recall of the static hanging string problem. This system has, obviously, been well studied in the literature (see, for example \cite{Behroozi, Fallis} for excellent discussions of the catenary as well as some of its variants) and we include it here only to establish our 
notations, discuss various parameterizations and benchmark our numerical integration scheme. So, consider
a string of constant length $2\ell_0$ and linear mass density
$\mu$ suspended in a gravitational field (see fig.~\ref{fig1}). Its shape 
can be obtained by minimizing its action which, for a static configuration reduces to its potential energy, 
\begin{equation}\label{stat1}
S=\int L\, d\xi,\qquad
L= g\mu\sqrt{X^{\prime2}+ Y^{\prime2}}\, Y
\end{equation}
where $\xi$ is a parameter along the string. Depending on the
choice of the coordinate $\xi$, there are at least two ways
to find the solution that extremises Eq.~(\ref{stat1}) to get the
shape of the string
\begin{equation}\label{stat1b}
x=X(\xi),\qquad
y=Y(\xi).
\end{equation}

\subsection{Choosing $\xi=x$}

In this case, we impose that $X=x$ in order to determine $y=Y(x)$ by extremizing
\begin{equation}\label{stat2}
S=\int L\, dx,\qquad
L= g\mu Y\sqrt{1+\left(\frac{dY}{dx}\right)^2}\,.
\end{equation}
First, note that since $L$ is independent of $t$ and $\dot Y$, 
the Hamiltonian  $H=L-\dot Y \partial L/\partial \dot Y$ reduces to $H=L$ so 
that mimnimizing  the action $S$ is equivalent to minimizing the energy.

Moreover, since $L$ is independent of $x$ also, we deduce that $L - Y' \partial L/\partial Y'$ is constant,
i.e.
\begin{equation}
 g\mu Y = k_x\sqrt{1+Y^{\prime2}},
\end{equation}
$k_x$ being a constant. This equation is easily integrated to give
\begin{equation}
 Y(x) = k \cosh\frac{x}{k} + C, \qquad k =\frac{k_x}{g\mu},
\end{equation}
the celebrated catenary. The boundary conditions $Y(\pm a)=y_A$ impose that $C=y_A-\ell \cosh(a/\ell)$ and the length of the string is found to be
$$
2\ell_0=\int_{-a}^{a} \sqrt{1+\left(\frac{dy}{dx}\right)^2} dx = 2 k \sinh\frac{a}{k}
$$
so that the full solution is
\begin{equation}\label{stat6}
 y=Y(x) = k\left[\cosh\frac{x}{k} - \cosh\frac{a}{k}\right]+y_A,
\end{equation}
with the constraint
\begin{equation}
 \ell_0 = k\sinh\frac{a}{k}.
\end{equation}
$k$ is indeed a complicated expression of the two parameters $(\ell_0,a)$ but the
solution is nevertheless unique and completely determined. In particular, 
we can compute the angle between the tangent to the string
and the $x$-axis as
\begin{equation}
 \tan\alpha = \frac{dY}{dx} = \sinh\frac{x}{k}.
\end{equation}

\subsection{Choosing $\xi=s$}

The previous argument assumes that one can choose $X=x$,
which is not {\it a priori} guaranteed, especially for non-static
configurations. An alternative choice is to use the length
along the string $s$ (defined through $dX^2+dY^2=ds^2$) as the parameter so that the length of the string is
\begin{equation}\label{c1}
 2\ell_0=\int_{-\ell_0}^{\ell_0}\sqrt{dX^{2}+ dY^{2}}ds=\int_{-\ell_0}^{\ell_0}ds.
\end{equation}
Since we need to extremize the action~(\ref{stat1}) subject to~(\ref{c1}),
we have to use a Lagrange multiplier to impose this constraint. Consequently, we
need to extremize
\begin{equation}\label{stat3}
S=\int  g\mu\sqrt{X^{\prime2}+ Y^{\prime2}}\, Y ds +
\lambda \left(\int\sqrt{X^{\prime2}+ Y^{\prime2}}ds-1\right).
\end{equation}
The variation with respect to $\lambda$, $X$ and $Y$ gives
the set of equations
\begin{eqnarray}
 && X^{\prime2}+ Y^{\prime2}=1,\label{stat11} \\
 &&  X' (g\mu Y+\lambda) = k g\mu,\label{stat12} \\
 && \left[\frac{Y'}{\sqrt{X^{\prime2}+ Y^{\prime2}}} (g\mu Y+\lambda)\right]' = g\mu.\label{stat13}
\end{eqnarray}
The first two of these can be combined  to obtained a closed
differential equation for $\hat Y(s)=Y+\lambda/g\mu$ as
\begin{equation}
 \frac{k^2}{\hat Y^2} + \hat Y^{\prime2}= 1,
\end{equation}
the solution of which is
\begin{equation}
 \hat Y(s) =\pm\sqrt{(s-s_0)^2 + k^2}.
\end{equation}
We can always choose $s=0$ at the middle of the string (the so-called affine parameterization)
so that $s\in[-\ell_0,\ell_0]$ and $X(s)$ can then be obtained by integrating
$X' \hat Y = k$ so that the full solution reads
\begin{eqnarray}
 && X(s) = k \sinh^{-1}\left( \frac{s}{k}\right) \label{e16}\\
 &&  Y(s) = \sqrt{s^2 + k^2}-\sqrt{\ell_0^2 + k^2}+y_A,\label{e17}
\end{eqnarray}
and $k$ is then fixed from the constraint  $X(\ell_0)=-X(-\ell_0)=a$ so that
\begin{equation}\label{eq-k}
 a = k \sinh^{-1}\left( \frac{\ell_0}{k}\right)\Longleftrightarrow \ell_0 = k\sinh \frac{a}{k}
\end{equation}
$\lambda$ is fixed by the boundary condition of $Y$
\begin{equation}
\frac{\lambda}{g\mu}=\sqrt{\ell_0^2+k^2}-y_A.
\end{equation}
Indeed, eliminating $s$ leads to the solution~(\ref{stat6}) found in the
previous paragraph. 

\begin{figure}[!htb]
\includegraphics[width=8cm,angle=0]{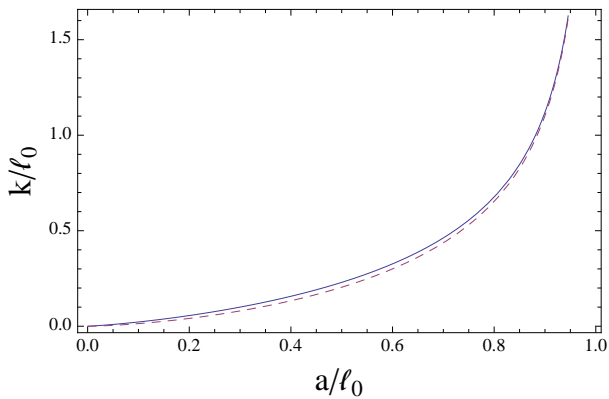}
\caption{The function $k(a)$. The dashed line represents
the asympotic expansion~(\ref{k-approx}) when $a/\ell_0\rightarrow1$.} 
\label{fig3}
\end{figure}

The function $k(a)$ is related to the auxilliary function $U(a)=k(a)/a$
that satisfies
\begin{equation}
 U \sinh \frac{1}{U} = \frac{\ell_0}{a}.
\end{equation}
Indeed, this equation has a solution only if $a/\ell_0<1$. Let us discuss the asymptotic behaviour
of $U$ and $k$.
\begin{itemize}
 \item When $U\rightarrow\infty$, $U \sinh \frac{1}{U} \rightarrow 1$, 
 from which we deduce that when $a\rightarrow \ell_0$, $U\rightarrow\infty$. Expanding
 $U \sinh \frac{1}{U}$ in a series in $1/U$, we get, to leading order, that $(6U^2)^{-1}=\ell_0/a-1$ 
 so that
 \begin{equation}\label{k-approx}
  k(a)\sim \frac{a}{\sqrt{6( \frac{a}{\ell_0}-1)}}
 \end{equation}
 \item When  $a\rightarrow 0$, $U\sinh\frac{1}{U}\rightarrow\infty$ so that $U\rightarrow0$
 and $k\rightarrow0$. 
 \end{itemize} 
The function $k$ is depicted in Fig.~\ref{fig3}. In this parametrisation, it follows from Eq.~(\ref{stat11}) that 
\begin{equation}
 X'(s)=\cos\alpha(s),\qquad  Y'(s)=\sin\alpha(s),
\end{equation}
where $\alpha(s)$ is the angle between the tangent vector to the string,
${\bf t}(s)=(X'(s),Y'(s))$ and the $x$-axis. It is then easily checked that 
\begin{eqnarray}
&& \tan\alpha(s) = Y'/X' = \frac{s}{k},\nonumber\\
&& \sin\alpha(s) = Y' = \frac{s/k}{\sqrt{1+s^2/k^2}},\nonumber\\
&& \cos\alpha(s) = X' =  \frac{1}{\sqrt{1+s^2/k^2}}.\label{stat22}
\end{eqnarray}

\subsection{Tension along the string}

\begin{figure}[!htb]
\includegraphics[width=8cm,angle=0]{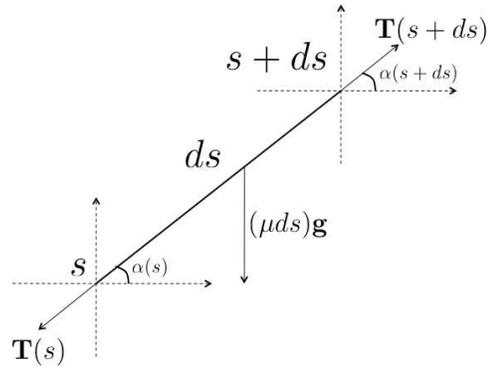}
\caption{A piece of string of length $ds$ is in equilibrium and subject to
the tension from the neighboring pieces and its weight.} 
\label{fig2}
\end{figure}

From the previous solution, one can derive the tension along the string,
${\bf T}=(T_x(s),T_y(s))=T(s){\bf t}(s)$. While tension, being
an internal force does not appear in the Lagrangian formulation, it can
nevertheless be defined from the Newton equations (see Fig.~\ref{fig2}) 
which, for
a static configuration, imply that
\begin{equation}\label{stat25}
T_x(s+ds)-T_x(s) = 0, \quad
T_y(s+ds)-T_y(s) = (\mu ds) g
\end{equation}
since the horizontal component is constant ({\it i.e.} the string experiences no tensive force in this direction) while
the vertical component has to balance the weight. It follows that
\begin{equation}
 (T_x)' = (T\cos\alpha)'=0,\quad
  (T_y)' = (T\sin\alpha)'=g\mu. 
\end{equation}
The second equation implies that $T_y=T(s)\sin\alpha(s) = \mu g s$
(since by symmetry $T_y(0)=0$). Using the expessions~(\ref{stat22}) we
conclude that $T_x(s)=T(s)\cos\alpha(s) =k\mu g$. We conclude
that 
\begin{equation}\label{eq-tension}
 T(s) = k\mu g \sqrt{1+\frac{s^2}{k^2}}=\mu g \hat Y(s).
\end{equation}

\subsection{Energy of the configuration}

\begin{figure}[!htb]
\includegraphics[width=8cm,angle=0]{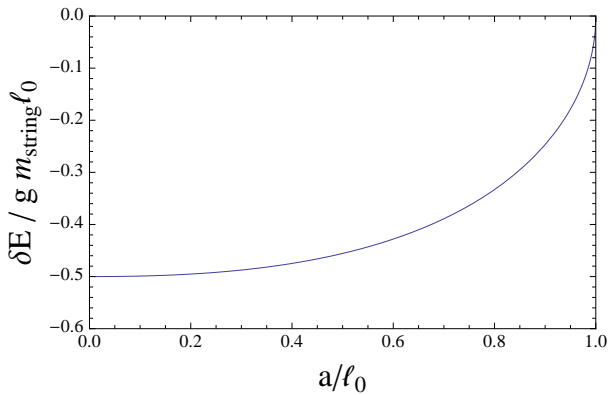}
\caption{Variation of $\delta E$, in units of $g m_{\rm string}\ell_0$,  as a function of $a$.} 
\label{fig4}
\end{figure}

Once the configuration of the string is known, we can compute its energy
as a function of the two parameters $(\ell_0,a)$ of the problem,
\begin{equation}
 E[a,\ell_0]= 2g\mu\int_0^{\ell_0} Y(s) ds,
\end{equation}
i.e.
$$
 \frac{E[a,\ell_0]}{2g\mu} =\ell_0\left(y_A-\sqrt{\ell_0^2+k^2}\right)+
 \int_0^{\ell_0}\sqrt{s^2+k^2} ds.
$$
The integral reduces to $\frac12 [s\sqrt{k^2+s^2}+k^2\ln(s+\sqrt{k^2+s^2})]^{\ell_0}_0$
so that, using  Eq.~(\ref{eq-k}) to express $\sinh^{-1}[\ell_0/k]$, one gets
$E[a,\ell_0]= g m_{\rm string} y_A + \delta E$ with $ m_{\rm string}=2\mu\ell_0$
and
\begin{equation}
 \frac{\delta E[a,\ell_0]}{g m_{\rm string}\ell_0} =-\frac{1}{2}\sqrt{1+ \frac{k^2}{\ell_0^2}}+
 \frac{k}{2\ell_0} \frac{a}{\ell_0}.
\end{equation}
In the limit $a\rightarrow\ell_0$, the string is stretched along $y=y_A$
and, as we have seen, $k/\ell_0\rightarrow\infty$ so that $\delta E\rightarrow 0$ and
$E\rightarrow g m_{\rm string} y_A$, as expected. In the
limit $a\rightarrow 0$, the string hangs vertically
and Eq.~(\ref{eq-k}) implies that $k\rightarrow 0$ so that
$E\rightarrow g m_{\rm string} (y_A-\ell_0/2)$, which is also
expected since the position of the barycenter of the string
is at $y_A-\ell_0/2$. Fig.~\ref{fig4} gives the dependence
of $E$ with $a$.

\section{The Dynamical catenary}

We now allow the string to move, keeping in mind the study of two
main situations: (1) the case where the points $A$ and $A'$ can move freely on
the $y=y_A$ axis and (2) the case of waves propagating along the static catenary.
It is clear from Fig.~\ref{fig4} that the string will tend to minimize its energy
so that we expect $a\rightarrow0$. The goal of this section is in particular
to describe this dynamics and to estimate the time scale for the two
end-points to meet (as a function of the parameters in the problem).

As in the previous section, we choose $\xi=s$ and start from the
action
\begin{eqnarray}
S&=&\int  \mu\sqrt{X^{\prime2}+ Y^{\prime2}}
           \left[\frac{1}{2}(\dot X^2+\dot Y^2) + g Y \right] ds dt \nonumber \\
           &&+\lambda \left(\int\sqrt{X^{\prime2}+ Y^{\prime2}}ds dt -1\right),\label{dyn1}
\end{eqnarray}
where the string is parameterized as $\left\{ X(s,t),Y(s,t)\right\}$.

\subsection{Equations of motion}

As for the static case, the equations of motion can be derived either
from Newton laws or by extremizing the action.

Let us first start from the Newton laws. Eq.~(\ref{stat25}) generalizes to
\begin{eqnarray}
 &&\mu ds \ddot X = T_x(s+ds)-T_x(s)=\partial_s T_x ds,\nonumber\\
 &&\mu ds \ddot Y = T_y(s+ds)-T_y(s)=\partial_s T_y ds - g\mu ds.\nonumber
\end{eqnarray}
Keeping in mind that $(X',Y')=(\cos\alpha,\sin\alpha)$, we end up
with the system for $\lbrace X(s,t),Y(s,t),T(s,t)\rbrace$
\begin{eqnarray}
 && \ddot X = \frac{T}{\mu} X'' + \left( \frac{T}{\mu}\right)' X',\label{ds1}\\
 &&\ddot Y = \frac{T}{\mu} Y'' + \left( \frac{T}{\mu}\right)' Y' - g,\label{ds2}\\
 && X^{\prime2}+ Y^{\prime2}=1,\label{ds3}
\end{eqnarray}
for $s\in]-\ell_0,\ell_0[$. It does not apply to $s=\pm\ell_0$ since there the
tension is discontinuous and this should be provided by the
boundary conditions.

Starting from the Lagrangian, we have to extremise~(\ref{dyn1}) with
respect to $X$, $Y$ and $\lambda$. This provides the three equations
\begin{eqnarray}
 && \ddot X(s,t) =\left[X'\left(g Y +\frac{1}{2}\dot X^2+\frac{1}{2}\dot Y^2+\frac{\lambda}{\mu}\right) \right]',\label{ds4}\\
 &&  \ddot Y(s,t) =\left[Y'\left(g Y +\frac{1}{2}\dot X^2+\frac{1}{2}\dot Y^2+\frac{\lambda}{\mu}\right) \right]'-g,\label{ds5}\\
 && X^{\prime2}(s,t)+ Y^{\prime2}(s,t)=1.\label{ds6}
\end{eqnarray}
Again, the tension does not appear in this derivation but it is clear that the systems~(\ref{ds4}-\ref{ds6})
and~(\ref{ds1}-\ref{ds3}) are equivalent if one sets
\begin{equation}\label{t00}
 T(s,t)=\mu\left(g Y +\frac{1}{2}\dot X^2+\frac{1}{2}\dot Y^2+\frac{\lambda}{\mu}\right).
\end{equation}

\subsection{Moving end-points}

\subsubsection{Full system of equations}

In order to solve the previous system, one should impose the boundary conditions.
We start by assuming that the end-points can move freely by remaining at
a consant height, that is
\begin{equation}
 Y(\pm\ell_0,t)=y_A, \qquad \dot Y(\pm\ell_0,t)=\ddot Y(\pm\ell_0,t)=0.\label{boundary}
\end{equation}
For the end-points, we define $X_A(t)=X(\ell_0,t)$ and $\alpha_0(t)=\alpha(\ell_0,t)$.
The tension being discontinuous (because this is the end of the string), we should have
$$
 \mu ds \ddot X_A = -T_x(\ell_0),\quad
 \mu ds \ddot Y _A= -T_y(\ell_0) - g\mu ds.
$$
The boundary conditions imply  that $T_y(\ell_0)= - g\mu ds$
and thus we conclude the motion of the end-point is dictated by\footnote{\label{fn5}
This condition can also be obtained from Eqs.~(\ref{ds4}-\ref{ds5}) since,
when evaluated in $s=\ell_0$ a derivative term of the form
$f'(\ell_0)$ can be evaluated as $f'(\ell_0)ds=f(\ell_0+ds)-f(\ell_0)=-f(\ell_0)$.}
\begin{equation}\label{e46}
  \ddot X_A = -g \cot \alpha_0(t).
\end{equation}

If we assume that the system is prepared in a static configuration and then
evolves freely, the final system should read
\begin{eqnarray}
 && \ddot X(s,t) = \left[\frac{T(s,t)}{\mu} X'(s,t)\right]',\label{fulldynamics1}\\
 &&\ddot Y (s,t)=  \left[\frac{T(s,t)}{\mu} Y'(s,t)\right]' - g,\\
 && X^{\prime2}(s,t)+ Y^{\prime2}(s,t)=1,\\
 && \ddot X(\ell_0,t) = -g \cot \alpha(\ell_0,t),\\
 && X(s,t=0) = x_{\rm stat}(s),\\
 && Y(s,t=0) = y_{\rm stat}(s),\\
 && T(s,t=0) = T_{\rm stat}(s).\label{fulldynamics2}
\end{eqnarray}

\subsubsection{Adiabatic solution}\label{sec-adiab}

In order to find an approximate solution to this system, let us assume
that the motion of the string is well approximated by a succession of static
configurations. This is a reasonable approximation as long
as the end-points are moving slowly.

Under such an approximation, $\cot \alpha(\ell_0,t)=\cot\alpha_{\rm stat}(t)=k(t)/\ell_0$
so that the motion of the end-points simplifies to
\begin{eqnarray}
 \ddot X_A(t) = -g \frac{k(t)}{\ell_0},\qquad
  \frac{\ell_0}{k(t)}=\sinh\frac{X_A(t)}{k(t)}.
\end{eqnarray}
Setting $X_A = a(t) = \ell_0 \underline{a}(t)$, $k(t)=\ell_0 K(t)$ and using the rescaled time $\tau = t/t_c$ with $t_c^2=\ell_0/g$, 
we have to solve
\begin{eqnarray}\label{e-init}
 \ddot {\underline{a}}(\tau) = -K[\underline{a}(\tau)],\qquad
  K(\tau)\sinh\frac{\underline{a}(\tau)}{K(\tau)}=1,
\end{eqnarray}
with a dot being $d/d\tau$.  

\begin{figure}[!htb]
\includegraphics[width=7.5cm,angle=0]{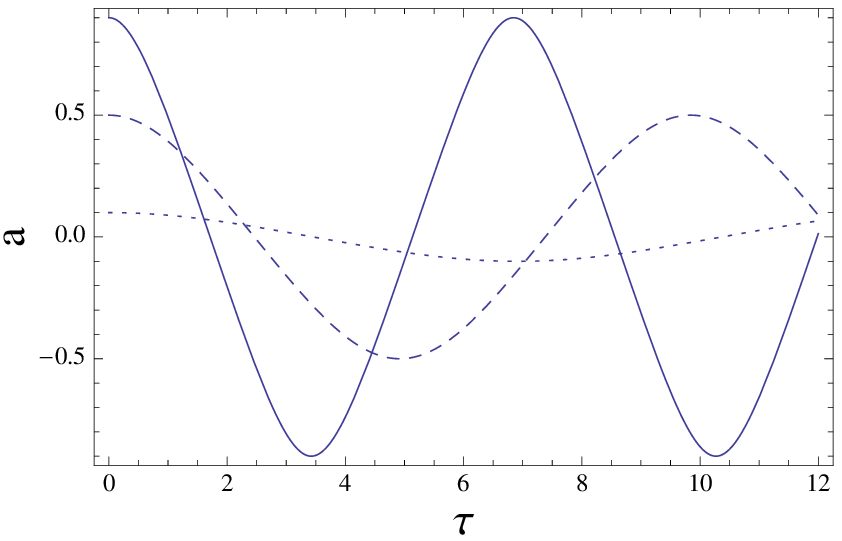}

$\quad$\includegraphics[width=8.7cm,angle=0]{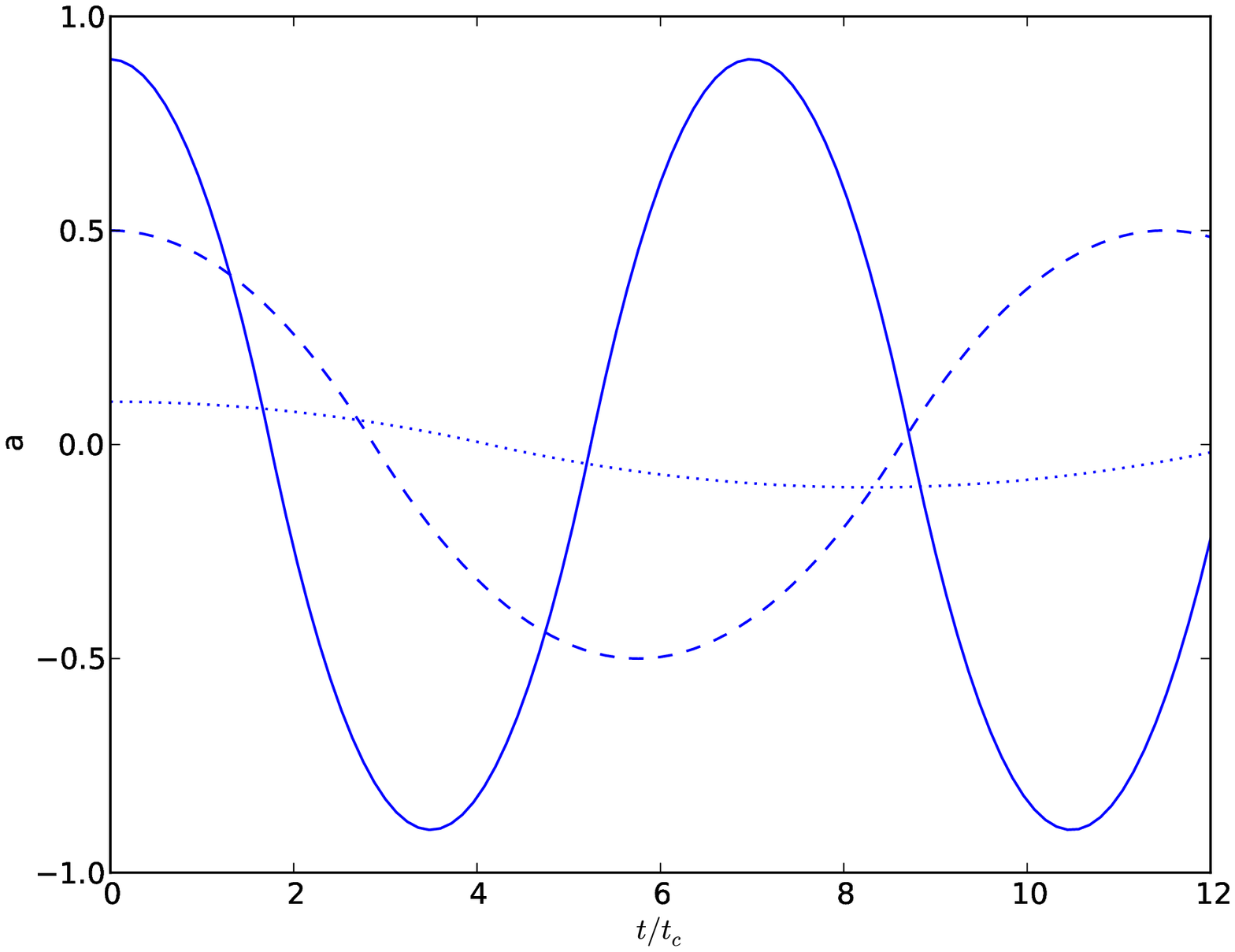}
\caption{Evolution of $a(\tau)$ in units of $\ell_0$ for $\underline{a}_{\rm in}=0.1$ (dotted line), 0.5 (dashed
line) and 0.9 (solid line) for the adiabatic solution (top)
and full numerical intergration (bottom). The frequency of the oscillation depends on $\underline{a}_{\rm in}$,
contrary to a standard harmonic oscillator. The agreement between the ywo computations is excellent,
at least for the first oscillations (see \S~\ref{subsecnum}).} 
\label{fig5}
\end{figure}

\begin{figure}[!htb]
\includegraphics[width=8cm,angle=0]{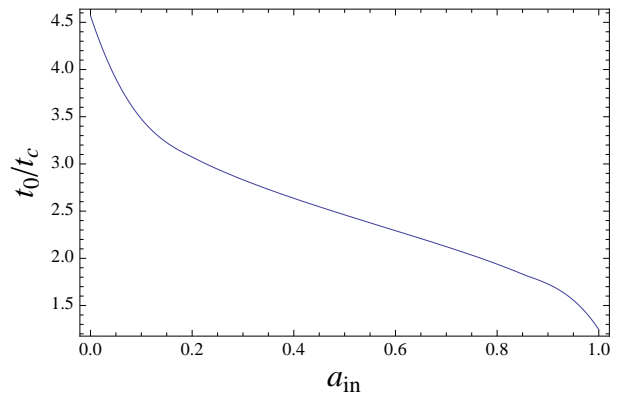}
\caption{Collapse time in units of $t_c=\sqrt{\ell_0/g}$ as a function of the initial amplitude $a_{\rm in}$.} 
\label{fig6}
\end{figure}

We can solve this equation numerically for different values of the initial amplitude
$a_{\rm in}$ (see Fig.~\ref{fig5}) and then compute the time $t_0$ for the
two end-points to meet (see Fig.~\ref{fig6}). For a standard harmonic oscillator,
$t_0$ is independent of the initial amplitude, contrary to our case at hand.

\subsubsection{Numerical integration}\label{subsecnum}

The system (\ref{fulldynamics1}-\ref{fulldynamics2}) can be solved numerically. However, to facilitate the numerical integration,  
it will prove more convenient to rewrite it as a system of first order equations of the form,
\begin{equation}
    \dot{X}(s,t) = u(s,t), \quad  \dot{u}(s,t) = \frac{d}{ds}F[X(s,t), X'(s,t)] .
\end{equation}

Differential equations of this form are of the general class that may be solved by either  
the Lax-Friedriechs finite difference~\cite{nr} or Smoothed Particle Hydrodynamics~\cite{sph} numerical integration schemes. To understand the exact behaviour of the dynamical catenary (and check the range of validity of the assumption of adiabaticity), we have implemented both of them and checked that they give the same results. 

\begin{figure}[!htb]
    \includegraphics[width=8cm,angle=0]{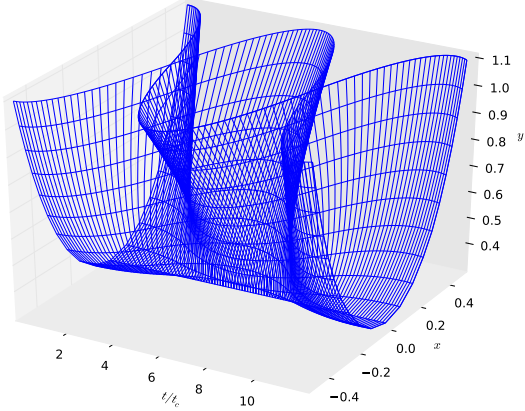}
    \caption{Numerical solution of the full string dynamics with $\underline{a}(0) = 0.5$.
    We see that the adiabatic solution is a good qualtitative description before the
    first crossing of the end-points. Thereafter, the development of cusps causes the adiabatic 
    approximation to break down.}
    \label{fig7}
\end{figure}

A surface plot of a solution is depicted in Fig.~\ref{fig7} and 
the motion of the endpoints for different configurations is shown in Fig.~\ref{fig5} (bottom)
for the case of a constant length string.
The first plot shows that the adiabatic solution is valid at most for the
first oscillations since after this, cusps developed and cannot be treated by the
adiabatic approximation.

\subsection{Waves}

As a second example, let us consider waves propagating on a static
catenary, so that we study small deviations and set
${\bf X}(s,t)={\bf X}_{\rm stat}(s) + {\bf u}(s,t)$ with
${\bf X}(s,t)=\lbrace X(s,t),Y(s,t),Z(s,t)\rbrace$, ${\bf X}_{\rm stat}=
\lbrace X_{\rm stat}(s),Y_{\rm stat}(s), 0 \rbrace$ and
${\bf u}(s,t)=\lbrace u_x(s,t),u_y(s,t),u_z(s,t)\rbrace$.

Let us start by neglecting the perturbation transverse to the plane of
the catenary, i.e. $u_z=0$. At first order in ${\bf u}$, the condition~(\ref{ds6})
implies that
\begin{equation}
 {\bf t}\cdot{\bf u}' = X'_{\rm stat}u_x' + Y'_{\rm stat}u_y' = 0.
\end{equation}
The perturbation can be chosen to be perpendicular to the string, which
implies that we must have ${\bf u}=f(s,t){\bf N}(s)$ with ${\bf N}(s)=(-Y_{\rm stat}',X_{\rm stat}',0)$
and $f(s,t)=  X'_{\rm stat}u_y  - Y'_{\rm stat}u_x$. Expanding Eqs.~(\ref{ds4}-\ref{ds5})
and then combining them, one can extract the equation of evolution of $f(s,t)$
\begin{equation}
 \ddot f -\frac{T_{\rm stat}}{\mu} f'' = \left(\frac{T_{\rm stat}}{\mu}\right)' f' \label{waves}
\end{equation}
where the tension along the string is given $T_{\rm stat}(s)$ is given in Eq.~(\ref{eq-tension}).
It describes damped waves with a sound speed
\begin{equation}
 c^2(s) = k g \sqrt{1+\frac{s^2}{k^2}}.
\end{equation}
The boundary conditions are $f(\pm\ell_0,t)=\dot f(\pm\ell_0,t)=f'(\pm\ell_0,t)=0$.
When $T_{\rm stat}\rightarrow{\rm cste.}$ we recover the standard wave equation
for the perturbation of a string.

Concerning the perturbations perpendicular to the plane of the catenary, one needs
to extend our action to
\begin{eqnarray}
S &=& \int \mu\sqrt{X^{\prime2}+ Y^{\prime2}+Z^{\prime2}}\, 
\left[\frac{1}{2}(\dot X^2+\dot Y^2+\dot Z^2) + g Y \right] ds dt \nonumber \\
&&+ \lambda \left(\int\sqrt{X^{\prime2}+ Y^{\prime2}+Z^{\prime2}}ds-1\right)\label{zpert}
\end{eqnarray}
in order to allow for a motion along the $z$-direction.
The variation with respect to $\lambda$ give the constraints~(\ref{ds6}) at zeroth and first
order in $u_z$. Then the variation with respect to $Z$, at first order in $u_z$ gives
(since it has no zeroth order)
\begin{equation}
 \ddot u_z -\frac{T_{\rm stat}}{\mu} u_z'' = \left(\frac{T_{\rm stat}}{\mu}\right)' u_z',
\end{equation}
that is the same equation as for the waves in the plane of the catenary.

\begin{figure}[!htb]
    \includegraphics[width=8cm,angle=0]{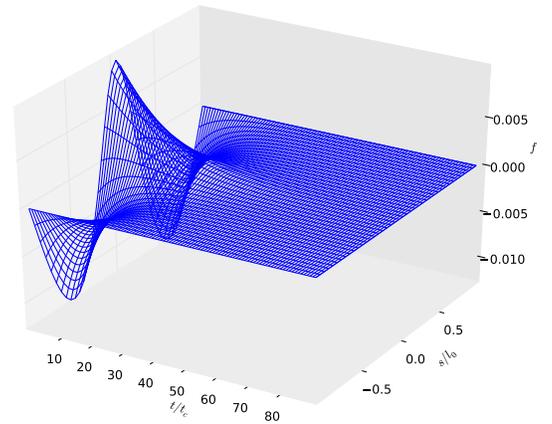}
    \caption{A numerical solution of $f(t,s)$ with $u_x(0,s)=
    A\sin(\pi s)$ and $u_y(0,s)=A\sin(1.5 \pi s)$.
    $A$ is fixed by the string length.
    The perturbations decay like an over damped harmonic oscillator.
    This can be seen in the figure as the bumps die out monotonically.}
    \label{fig10}
\end{figure}

\begin{figure}[!htb]
    \includegraphics[width=8cm,angle=0]{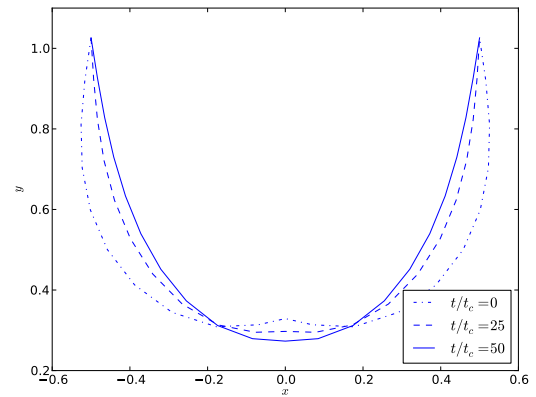}
    \caption{The string profile at different times using the same initial configuration as in 
    Fig.~\ref{fig10}. The string coordinates here where calculated using 
    $X = X_{\rm stat} - f Y^\prime$ and $Y = Y_{\rm stat} + f X^\prime$.}
     \label{fig11}
\end{figure}

Fig.~\ref{fig10} depicts the evolution of $f$ with time assuming an initial profile with $u_x(s,t)=A\sin(\pi s)$ and $u_y(s,t)=A\sin(1.5 \pi s)$.
$A$ is chosen so that the perturbed string's length is the same as the static sting.
Using the relation ${\bf u}(s,t)=f(s,t){\bf N}(s)$, we can calculate the perturbations at later time steps.
Fig.~\ref{fig11} shows the string profile at different times.

\section{Relativistic extension}

A string is a (1+1)-dimensional object propagating in a
(1+3)-dimensional spacetime with metric $g_{\mu\nu}$
with $\mu,\nu=0\ldots3$. It is thus described by four
embedding function $X^\mu(\xi^a)$ where $\xi^a$ are
(2-dimensional) intrinsic coordinates on the worldsheet
($a,b=0,1$).

The local geometry of the string worldsheet is thus determined
by the metric $\gamma_{ab}$ induced by the spacetime metric $g_{\mu\nu}$
on the string worldsheet and is given by
\begin{equation}
 \gamma_{ab} = g_{\mu\nu} X^\mu_{,a} X^\nu_{,b}
\end{equation}
where $X^\mu_{,a}\equiv \partial_a X^\mu$.

Note that a 2-dimensional surface being conformally flat, it always exist 
a system of coordinates such that $\gamma_{ab}=\Omega(\xi^c)
\eta_{ab}$, $\eta_{ab}$ being the 2-dimensional Minkowski metric~\cite{conf}.

\subsection{General dynamics}

The relativistic extension of the dynamics of a string derives from the Goto-Nambu
action~\cite{gn1,gn2}. which states that the action is proportional to the
area of the 2-dimensional worldsheet spanned by the string,
\begin{equation}\label{relat1}
 S = -T\int \sqrt{-\gamma}d^2\xi,
\end{equation}
where $\gamma$ is the determinant of $\gamma_{ab}$.
Interestingly, extremising this action with respect to the 
2-dimensional metric gives the general equation
of motion~\cite{Carter:2000wv}
\begin{equation}\label{relat2}
 \frac{1}{\sqrt{-\gamma}}\partial_a[\sqrt{-\gamma}\gamma^{ab}\partial_b
 X^\mu] + \Gamma^\mu_{\nu\rho}\gamma^{ab}\partial_a X^\nu\partial_b X^\rho = 0,
\end{equation}
where $\Gamma^\mu_{\nu\rho}$ are the Christoffel symbols of the
background spacetime,
$$
 \Gamma^\mu_{\nu\rho} = g^{\mu\lambda}
 \left(\partial_\mu g_{\lambda\rho} + \partial_\rho g_{\mu\lambda} -\partial_\lambda g_{\mu\rho}\right).
$$
Eq.~(\ref{relat2}) is a generalization of the geodesic equation of a point particle
(for which the worldsheet is 1-dimensional) to higher-dimensional objects.

This was extensively used to study the dynamics of cosmic strings (see
Ref.~\cite{kibble} for a review). In Minkowsky space, the conformal
gauge leaves considerable freedom and one can further
impose that $X^0=t$, known as the {\em temporal gauge}.
The gauge conditions and the evolution equations then takes the form
\begin{equation}
 \delta_{ij}\dot X^i \dot X^j =1, \quad
  \delta_{ij}\dot X^i  X^{\prime j} =0,\quad
 \ddot X^i - X^{\prime\prime i}=0. 
\end{equation}

\subsection{Relativitic catenary}

In order to describe a relativistic accelerated string, we consider that
the string is embedded in a Rindler spacetime with metric~\cite{Berenstein-Chung}
\begin{equation}
 g_{\mu\nu} dx^\mu dx^\nu = -\frac{\kappa^2}{c^2} y^2 dt^2 + dx^2 + dy^2,
\end{equation}
where we have dropped the third spatial dimension since it is irrelevant
in our problem. This spacetime describes part a wedge of the Minkowski
spacetime since the change of coordinates\footnote{
These coordinates were introduced by A. Einstein and N. Rosen in 1935 
and popularized by W. Rindler in 1956 in his book~\cite{rindlerbook}.} 
\begin{equation}\label{rindmet}
 c\eta = y\sinh \frac{\kappa t}{c}, \qquad
 v = y\cosh \frac{\kappa t}{c},\qquad
 u = x
\end{equation}
brings the metric~(\ref{rindmet}) to 
\begin{equation}
 g_{\mu\nu} dx^\mu dx^\nu =- c^2d\eta^2 + du^2 + dv^2.
\end{equation}
This choice of embedding spacetime can be understood from the dynamics of
the relativistic motion of a point particle subject to a constant acceleration.
$u^\mu$ being the tangent vector to its worldline $x^\mu(\tau)$, such that
$u^\mu u_\mu =-c^2$, the 4-acceleration is defined as $\gamma^\mu=d u^\mu/d\tau$,
$\tau$ being the proper time of the accelerated observer.
It is perpendicular to $u^\mu$ ($u^\mu\gamma_\mu=0$) and we assume it is
constant, i.e. $\gamma_\mu\gamma^\mu = g^2$. If we align the axis $v$ in the direction of the acceleration,
then $-u_0^2+u_1^2=-c^2$ and $-\gamma_0^2+\gamma_1^2=g^2$. The first equation
implies that $u_0= c\cosh f(\tau)$ and $u_1=c\sinh f(\tau)$ and the second equation gives that
$f(\tau)= g\tau/c$. The equation of the trajectory is then obtained as
$\eta(\tau)=\frac{c}{g}\sinh(g\tau/c)$ and $v(\tau)=\frac{c^2}{g}\cosh(g\tau/c)+h$, i.e.
$(v-h)^2-c^2\eta^2=c^4/g^2$.
It is then clear that a particle at a fixed valur of $y$ in the Rindler space; $y=y_A$
has wordline $v^2-c^2\eta^2=y_A^2$ which corresponds, as we have seen in the
previous paragraph, to the wordline of an observer uniformly accelerated in Minkowky.
At early times, one has
\begin{equation}
 v=y_A+\frac{\eta^2}{2y_A},
\end{equation}
which corresponds to the usual Newtonian expression if
\begin{equation}
\kappa=g=\frac{c^2}{y_A}
\end{equation}
Note that the metric is singular on $y=0$, often called the Rindler horizon and that
$g_{00}$ depends only on $y$, so that it describes a gravitational potential
attracting the particle towards $y=0$.

The worldsheet of the string is described by $X^\mu=\lbrace T,X,Y \rbrace$
and we decide to choose the intrinsic coordinates such that
$\xi_1=s$ and $\xi_0=t$. This latter choice of identifying the
time coordinate of the string worldsheet and of the background spacetime is usually 
refered to as the {\em temporal gauge}. Because of this choice, we cannot, in
general, choose coordinates such that $\gamma_{ab}$ is explicitely
conformal to $\eta_{ab}$. However, we can still choose $\xi_1=s$ so that
\begin{equation}\label{relat4}
  g_{\mu\nu} X^{\prime \mu}(s,t) X^{\prime \nu}(s,t) = 1
\end{equation}
and still have the freedom to impose
\begin{equation}\label{relat5}
  g_{\mu\nu} \dot X^\mu(s,t) X^{\prime \nu}(s,t) = 0.
\end{equation}
Concerning the boundary conditions, we assume that the string
hangs on a brane that seats at a constant distance from the
Rindler horizon so that it undergoes a constant acceleration. Note however
that the string does not lie at a constant distance from the horizon so
that the local acceleration (or gravitational potential) will vary
along the string, which is a major difference compared to the
Newtonian case.

\subsubsection{Equation of motion}

The equation of motion are obtained directly from the action~(\ref{relat1}) using that
$$
\gamma=\left(-\frac{\kappa^2}{c^2}Y^2+\dot X^2 + \dot Y^2\right)(X^{\prime2}+Y^{\prime2})
-\frac{1}{4}(\dot XX' + \dot YY')^2.
$$
The Euler equations take the form
\begin{eqnarray}
 &&\frac{\delta L}{\delta X} = \partial_s\left(\frac{\delta L}{\delta X'}\right) + \partial_t\left(\frac{\delta L}{\delta \dot X}\right)\\
 &&\frac{\delta L}{\delta Y} = \partial_s\left(\frac{\delta L}{\delta Y'}\right) + \partial_t\left(\frac{\delta L}{\delta \dot Y}\right)
\end{eqnarray}
reduce to the set of equations
\begin{eqnarray}
 &&\left(\frac{\dot X}{\sqrt{-\gamma}}\right)^. +\left[X' \sqrt{-\gamma} \right]'=
 0,\label{e66}\\
 &&\left(\frac{\dot Y}{\sqrt{-\gamma}}\right)^. +\left[Y' \sqrt{-\gamma} \right]'=
 \frac{\kappa^2}{c^2\sqrt{-\gamma}}Y,\label{e67}
\end{eqnarray}
once the constraints 
\begin{eqnarray}
 &&\dot X X' + \dot Y Y' = 0,\quad
 X^{\prime2}+Y^{\prime2} =1
\end{eqnarray}
are taken into account. It can be checked that these
equations are equivalent to Eq.~(\ref{relat2}).

\subsubsection{Static case}

For a static configuration, the previous set of equations reduces
to
\begin{equation}
 (X' Y)' = 0,\qquad
 (Y' Y)' = 1.
\end{equation}
We recognize the set~(\ref{stat11}-\ref{stat13}) and it is clear that the solution if again
a catenary. It follows that the static solution of a relativistic 
Goto-Nambu string in a Rindler spacetime is similar to a constant
length Newtonian catenary in an external gravitational field,
that is by Eqs.~(\ref{e16}-\ref{e17}). This is a non-trivial equivalence
since it tells that the configuration of a constant tension relativistic string
hanging in a Rindler space, and thus such that the gravitational
acceleration depends on height, is equivalent to a Newtonian
string of constant length and constant lineic mass hanging in
a gravitational potential with constant gravitational acceleration.

\subsubsection{Dynamical case}

The previous paragraph confirms the analysis
of Ref.~\cite{Berenstein-Chung} that we can now extend to the dynamical case.
Eqs.~(\ref{e66}-\ref{e67}) take the form
\begin{eqnarray}
&&\ddot X + f(X,Y) \dot X = -\sqrt{-\gamma}(X'\sqrt{-\gamma})'\label{dr1}\\
&&\ddot Y + f(X,Y) \dot Y = -\sqrt{-\gamma}(Y'\sqrt{-\gamma})' + \frac{\kappa^2}{c^2} Y\label{dr2}
\end{eqnarray}
where $f$ is a friction term given by
\begin{equation}
 f(X,Y) = -\frac{\dd\ln\sqrt{-\gamma}}{\dd t}.
\end{equation}
These equations are different from the Newtonian version in particular because
the gravitational potential is not constant along the string since it does not
lie at a $y=$~const. position.  However,
thanks to the fact that the static solution is the same as in the Newtonian
case, we can use the adiabatic solution to investigate
the motion of the end-points, as in \S~\ref{sec-adiab}. Eq.~(\ref{dr1}) implies
that 
$$
 \ddot X_A(t) +f \dot X_A + \sqrt{-\gamma}(X'\sqrt{-\gamma})'_{s=\ell_0}=0.
$$
The $s$-derivative at the end-point can be evaluated as in footnote~\ref{fn5},
using Eq.~(\ref{dr2}) which implies that
$ \sqrt{-\gamma}(Y'\sqrt{-\gamma})'_{s=\ell_0} =  \kappa^2y_A/c^2$.
Defining $\alpha$ as in Eq.~(\ref{stat22}), we conclude that
$\sqrt{-\gamma}(X'\sqrt{-\gamma})'_{s=\ell_0}=(\kappa^2y_a/c^2)\cot\alpha_0(t)
=g\cot\alpha_0(t)$. Then, the friction coefficient is obtain from $\gamma(s=\pm\ell_0)
=-(\kappa^2 y_A^2/c^2)+\dot X_A^2=-c^2+\dot X_A^2$ so that
$$
 f(\ell_0,t)=\frac{\dot X_A\ddot X_A}{c^2-\dot X_A^2}.
$$
The equation of motion of the end-point is thus
\begin{eqnarray}
 \frac{\ddot X_A(t)}{1-\frac{\dot X_A^2}{c^2}}= - g\cot\alpha(\ell_0,t).
\end{eqnarray}
We see that what appeared as a friction term recombines in order
to give an equation which is the pure relativistic invariant of
Eq.~(\ref{e46}). The Newtonian limit is recovered when
$\dot X_A^2/c^2\ll1$.

The adiabatic approximation assumes that
$\cot \alpha(\ell_0,t)=\cot\alpha_{\rm stat}(t)=k(t)/\ell_0$ (since the
static solution is identical to the Newtonian case) so that
the final system takes the form
\begin{eqnarray}
 &&\frac{\ddot X_A(t)}{1-\frac{\dot X_A^2}{c^2}}= - g\cot\alpha_{\rm stat}(\ell_0,t)= -g \frac{k(t)}{\ell_0},\\
 &&\frac{\ell_0}{k(t)}=\sinh\frac{X_A(t)}{k(t)},
\end{eqnarray}
once we have used the notation of \S~\ref{sec-adiab}, it reduces to
\begin{eqnarray}\label{e-init}
 \frac{\ddot {\underline{a}}(\tau)}{1-\frac{\ell_0}{y_A}\dot{\underline{a}}^2} = -K[\underline{a}(\tau)],\qquad
  K(\tau)\sinh\frac{\underline{a}(\tau)}{K(\tau)}=1,
\end{eqnarray}
with a dot being $d/d\tau$. We thus expect differences from the Newtonian
analysis when $\ell_0/y_A\not\ll1$, that is when the brane is
seating close enough from the Rindler horizon.

\begin{figure}[!htb]
\includegraphics[width=8cm,angle=0]{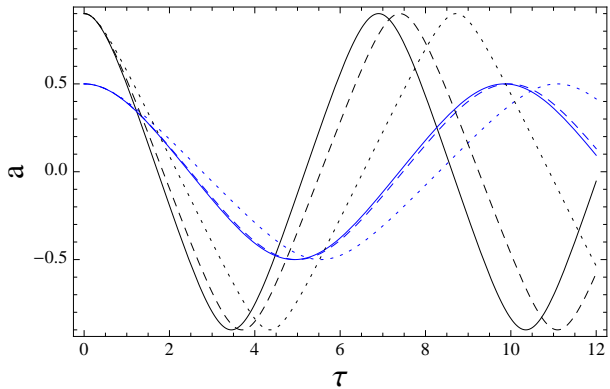}
\caption{Evolution of $a(\tau)$ in units of $\ell_0$ for $\underline{a}_{\rm in}=0.5$ (blue)
0.9 (black), respectively for $y_A/\ell_0=.1,1,10$ (dotted, dashed, solid) and 
$y_A/\ell_0=.3,1,10$ (dotted, dashed, solid).} 
\label{fig5bis}
\end{figure}

We can integrate this equation in the same way as the Newtonian case
and the solution is depicted on Fig.~\ref{fig5bis}. We deduce the typical
collapse time of the string in function of the initial separation of the end-points
and the distance of the D-brane from the Rindler horizon. By comparing
to Fig.~\ref{fig5}, we conclude that the order of magnitude obtained from
the Newtonian case is usually a good estimate. Indeed, when $\ell_0/y_A\ll1$,
the Newtonian description starts to fail.

\begin{figure}[!htb]
\includegraphics[width=8cm,angle=0]{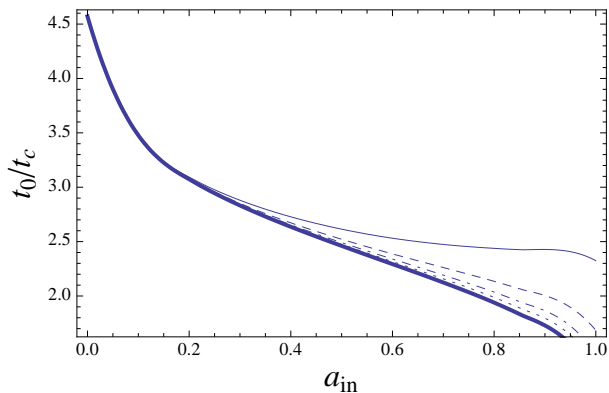}
\caption{Collapse time in units of $t_c=\sqrt{\ell_0/g}=\sqrt{\ell_0y_A}/c$ as a function 
of the initial amplitude $a_{\rm in}$ for $y_A/\ell_0=0.2,0.5,1,2,10$ (solid, dahed, dot-dashed, dotted,
thick).} 
\label{fig6bis}
\end{figure}

\section{Discussion}

The problem of a static string suspended in a gravitational field is an old one whose solution - the catenary - is well know to most freshman students of physics. In this article, we revisit this system with the intent to allow the (attached) ends of the string to move freely along the direction of suspension ({\it i.e.} along a fixed height). We have argued that
this simple mechanical system actually gives a remarkably good description of the dynamics of a 
long open string attached to an accelerated D-brane in string theory. This latter problem has an equivalent formulation as a string attached to a D-brane embedded in a Rindler spacetime at a fixed distance from the Rindler horizon.

In the Newtonian case, which forms the bulk of this article, we have obtained the typical collision time
for the endpoints of the string as a function of the relevant parameters and showed that it is typically of the order of $(1-4)\times\sqrt{\ell_0/g}$. This estimate, as well as a solution for the (rescaled) position of the string endpoint, were obtained through the assumption that the evolution of the string is adiabatic and checked by a direct numerical integration of the equations of motion. The numerical integration also showed that, in the absence of interactions between the string endpoints, the natural evolution of the catenary profile degenerates into cusps after one or two crossings.

The analysis was then extended to a relativistic string
of constant tension, {\it i.e.} of the Goto-Nambu type. Here we were able to gain some physical intuition why the static configuration of such a string attached to a D-brane located
in a Rindler spacetime, {\it i.e.} subject to a constant acceleration, was similar
to the constant length Newtonian catenary as found in \cite{Berenstein-Chung}. 
We also found that, while the dynamical relativistic case differs quite substantially from the 
Newtonian string, at least at the level of the equations of motion and their solutions, predictions for the collision-time for the string ends are remarkably similar, as long as the D-brane is not too close to the
Rindler horizon. Indeed, this is one of the main results of this analysis. In fact, this simple classical analogue gives us a number of important clues about the physics of radiating branes. For example, when string interactions are turned on, there are two ways that the long open string attached to the D-brane can form a closed string and leave the brane: either the endpoints of the string meet or the string self-intersects at some point in its evolution forming in the process a closed string loop and a shorter open string. From our analysis we see that the string does indeed self-intersect, forming several cusps along its length but that this only occurs {\it after} the endpoints coalesce. In its broadest sense, our results offer a way to construct a toy model of the dynamics of the process of gravitational radiation from accelerated branes. We describe this process in more detail in a companion article~\cite{catenary2}.

\section{Acknowledgements}
This work is based upon research supported by the National Research Foundation (NRF) of South Africa's Thuthuka (UID 61699) and Key International Scientific Collaboration programs (UID 69813).


\begin{thebibliography}{99}

\bibitem{general}
 H. Goldstein, C. Poole, and J. Safko, {\em Classical Mechanics} (Addison-Wessley, San-Francisco);
 V.I. Arnold, {\em Mathematical methods of classical Mechanics} (Springer-Verlag,
 New-York, 1989);
 
\bibitem{variation}
 R. Courant, and D. Hilbert,
 {\em Methods of Mathematical physics}
 (Wiley, New York, 1989), pp. 172 and 218. 

\bibitem{Behroozi}
F. Behroozi, P. Mohazzabi, and J.P. McCrickard,
``{\it Remarkable shapes of a catenary under the effect of gravity and surface tension}",
Am. J. Phys. {\bf 62}, 12, (1994)

\bibitem{Fallis}
M.C. Fallis,
``{\it Hanging shapes of nonuniform cables}",
Am. J. Phys. {\bf 65}, 2, (1997)

\bibitem{conf}
 M.B. Green, J. Schwartz, and E. Witten,
 {\it Superstring Theory} 
 (Cambridge: Cambridge University Press,1987).

\bibitem{Berenstein-et-al}
 D. Berenstein, D. Correa, and S. V$\grave{\textrm{a}}$zquez, 
``{\it A study of open strings on giant gravitons, spin chains and
integrability}," 
J. High Energy Phys. \textbf{065} 0609 (2006) 
\texttt{hep-th/0604123}

\bibitem{Berenstein-Chung}
D. Berenstein, and H.-J. Chung, 
``{\it Aspects of open strings in Rindler Space},"  
\texttt{arXiv:0705.3110 [hep-th]}

\bibitem{nr}
  W.H. Press, S.A. Teukolsky, W.T. Vetterling, and B.P. Flannery, 
  {\it Numerical Recipes: The Art of Scientific Computing, Third Edition} 
  (Cambridge: Cambridge University Press, 2007).

\bibitem{sph}
  G.R. Liu, and M.B. Liu, 
  {\it Smoothed Particle Hydrodynamics: A Meshfree Particle Method} 
  (Singapore: World Scientific Publishing Company, 2003)

\bibitem{gn1}
 Y. Nambu,  
 ``{\it Symmetries and Quark Models}'', 
 R. Chand  Eds. (New York Gordon and Breach, 1970).

\bibitem{gn2}
 P. Goddard, J. Goldstone, C. Rebbi,  and C. Thome,
  Nucl. Phys. {\bf 568} 109 (1973).
  
\bibitem{kibble}
 M.B. Hindmarsh, and T.W.B. Kibble, 
 ``{\it Cosmic strings}'',
 Rep. Prog. Phys. {\bf58}, 411-562 (1995) .
 
\bibitem{Carter:2000wv}
  B.~Carter,
  ``{\it Essentials of classical brane dynamics}'',
  Int.\ J.\ Theor.\ Phys.\  {\bf 40}, 2099-2130 (2001).
  [gr-qc/0012036].
  
\bibitem{rindlerbook}
 W. Rindler,
 {\it Special Relativity} (Oxford Science Publications, 1991).  
 
\bibitem{catenary2}
 D. de Klerk, J. Murugan, and J.-P. Uzan, 
 ``{\it A Toy Model of Gravitational Radiation from Accelerated D-branes}'',
 in preparation.
  
\end{thebibliography}
\end{document}